\documentclass[10pt,aps,prm,amsmath,amssymb,twocolumn,%
				showpacs,
				balancelastpage,raggedbottom, 
				superscriptaddress,
				citeautoscript,
				reprint,shortbibliography,%
				floatfix]{revtex4-2}
\pdfoutput=1
\usepackage{graphicx}
\usepackage{dcolumn}
\usepackage{bm}
\usepackage{ mathrsfs }
\usepackage{soul}
\usepackage[scaled=0.90]{helvet} 
\usepackage[protrusion=true,expansion=true,final]{microtype}
\usepackage{units,xspace}
\usepackage[usenames,dvipsnames]{xcolor}
\usepackage[bookmarks=false,colorlinks]{hyperref}
\usepackage{natbib}
\usepackage{textcase}
\usepackage{bm}
\usepackage[T1]{fontenc}
\usepackage{mathptmx}
\hypersetup{
    linkcolor=magenta,        
    citecolor=OliveGreen,     
    filecolor=Plum,      	
    urlcolor=MidnightBlue,           
}
\usepackage{graphicx}
\usepackage{dcolumn}
\usepackage{bm}
\usepackage{soul}
\usepackage{multirow}

\begin{document}

\title{Efficient Estimation of Material Property Curves and Surfaces \emph{via} Active Learning }
\author{Yuan Tian}
\author{Dezhen Xue}
\email{xuedezhen@xjtu.edu.cn}
\author{Ruihao Yuan}
\author{Yumei Zhou}
\email{zhouyumei@xjtu.edu.cn}
\author{Xiangdong Ding}
\author{Jun Sun}
\affiliation{State Key Laboratory for Mechanical Behavior of Materials, Xi'an Jiaotong University, Xi'an 710049, China}
\author{Turab Lookman}
\email{turablookman@gmail.com}
\affiliation{Los Alamos National Laboratory, Los Alamos, NM 87545,USA}

\date{\today}

\begin{abstract}

The relationship between material properties and independent variables such as temperature, external field or time, is usually represented by a curve  or surface in a multi-dimensional space.
Determining such a curve or surface requires a series of experiments or calculations which are often time and cost consuming.
A general strategy uses an appropriate utility function to sample the space to recommend the next optimal experiment or calculation within an active learning loop. 
However, knowing what the optimal sampling strategy to use to minimize the number of experiments is an outstanding problem.
We compare a number of strategies based on directed exploration on several materials problems of varying complexity using a Kriging based model.
These include one dimensional curves such as  the fatigue life curve for  304L stainless steel and the Liquidus line of the Fe-C phase diagram, surfaces such as the Hartmann 3 function in 3D space and the fitted intermolecular potential for Ar-SH, and a four dimensional data set of experimental measurements for BaTiO$_3$ based ceramics. We also consider the effects of experimental noise on the Hartmann 3 function. We find that directed exploration guided by maximum variance provides better performance overall, converging faster across several data sets. However, for certain problems, the trade-off methods incorporating exploitation can perform at least as well, if not better than maximum variance.
Thus, we discuss how the choice of the utility function depends on the distribution of the data, the model performance and uncertainties,  additive noise as well as the budget.

\end{abstract}

\maketitle

\section{Introduction}
The accurate prediction of the properties of materials as a function of independent variables is crucially important in exploiting their use in different applications. 
Such a functional relationship is usually described as a curve or surface between a property and the independent variable/s in a multidimensional  diagram. \cite{askeland2013essentials}
These properties can be mechanical, thermal, electrical, magnetic, optical, and chemical ones, and the independent variables usually include chemical composition, temperature, time and heat treatment conditions. \cite{newnham2005properties}
The materials property curves and surfaces can determine the variation behavior, critical states and property optima, and consequently play a crucial role in the design of new materials, the assessment of hazards and the optimization of processing parameters. 
Familiar examples include phase boundaries and surfaces in temperature versus composition space,  fatigue life cycle curves describing the relationship between mechanical properties and loading cycles, and intermolecular potential energy surfaces for molecules.

Determining  a property curve or surface is often time and cost consuming as a number of measurements or calculations are required depending on the accuracy needed.  
An adequate number of data needs to be accumulated as the independent variable is varied in given steps.  
The data requirements are sensitive to nonlinearities and sharp changes in the functional form as well as the presence and number of multiple extrema, including critical points.
For example, establishing a phase diagram requires a series of experiments to determine the critical temperature for different compositions or pressures. Similarly,  a number of parallel samples, each of which is used to obtain the ultimate stress for a given number of loading cycles, are required  to obtain the fatigue curve of an alloy,
This  therefore poses the challenge of accurately estimating the property curve and surface with as few measurements or calculations as possible.
Although regression algorithms have been employed to model the functional form between the property and the independent variables\cite{Shen2019Physical,Im2019Identifying,Stanev2018Machine, SHIN2019Modern},
the regressed results inevitably contain large uncertainties if the number of initial data points is relatively small, especially if the relationship between the property and independent variable is complex.
As the number of initial data points increases, more experiments or calculations are needed with concomitant increases in time and resources used.
Hence, there is a need for an approach that can predict general property curves/surfaces and successively refine them rapidly using as few  new measurements or calculations as possible.

Active learning or optimal experimental design allows an algorithm to choose the data from which it learns so that it may learn more efficiently with less training data than otherwise.\cite{Lookman2019Active, Zhang2019Active,Faneaay5063,Brown,Tran2018Active, Gubernatis2018Machine} 
This becomes particularly important in areas such as materials science where the size of a good quality labelled data set for supervised learning is often limited because of the expense associated with generating it. \cite{Casciato2012Optimization,Smith2018Less,Lookman2019Active,ramprasad2017machine}
It then becomes desirable to recommend one or more unlabelled instances from a large pool of possibilities to be labelled by experiments or calculations using active learning methods.\cite{Fukazawa2019Bayesian,Talapatra2019Experiment,Vasudevan2019Materials}
The goals are to achieve higher accuracy of prediction or exploit the optimum, with minimization of the overall cost for obtaining labeled data. \cite{Reker2015Active, Kapoor2010Gaussian,Kiyohara2016Acceleration, Murray1997Active, Schmidt2019Recent, Seko2015Prediction}
It is being increasingly applied to materials science data to efficiently guide and minimize the number of experiments needed. 
\cite{Ling2017High,Bassman2018Active, Wen2019Machine,Yuan2019Accelerated, xue2016accelerated,xue2016Acceleratedpnas,Terayama2019Efficient}

By invoking concepts from decision theory, various utility functions, which are akin to query strategies in active learning, can be defined to decide which instances of the unlabeled data would be most informative to be labeled. 
Sampling the most important states is therefore a problem of considerable importance to avoid excessive numbers of iterations or experiments, especially when one may not know which states are most important. This requires exploring the allowed space efficiently and well enough and the problem has been studied in the context of reinforcement learning. Exploration techniques are essentially of two kinds, undirected exploration and directed exploration \cite{Thrun1992Efficient}. 
Undirected exploration is uninformed and characterized by selecting actions randomly from a given distribution. If the distribution is uniform, we have random exploration in which costs or rewards are not taken into account. 
Whitehead \cite{Whitehead1991ASO} has proved that under certain conditions  random walk exploration leads to the result that the learning time scales exponentially with the size of the state space. In spite of this, undirected exploration is studied widely in the literature via numerical experiments as often the conditions leading to the analytical result are not fully satisfied. 

Directed exploration, on the other hand, uses knowledge to guide the exploration search so that the exploration rule directly determines which action to take next. The goal is to select actions which maximize the improvement over time, which is impossible to determine as we do not know in advance how a given decision will improve the performance. Thus, all directed exploration techniques are heuristics designed to optimize knowledge gain. 
 The exploration may be achieved by choosing states based on frequency of occurrence (Counter based exploration), or assumed to have a high prediction error (maximum variance), or those that include different degrees of exploitation functions based on using the best value of the model predictions at the time. Examples of the latter include trade-off methods such as the efficient global optimization (EGO) and knowledge gradient (KG) schemes based on Bayesian optimization for finding maxima/minima of functions. \cite{Bisbo2019Efficient,Theiler2017Selecting} All have the aim to optimize both learning time and learning costs. One of the few analytical results is that for Counter based exploration. Thurn \cite{Thrun1992Efficient} has shown that the worst-case complexity of learning, under given conditions, is always polynomial in the size of the state space.  

How much exploration needs to be performed depends on the costs of collecting new information and the value associated with that information. 
 In the absence of analytical results for realistic problems and strategies, what we are left with is a study of different heuristics with different degrees of interplay between exploration and exploitation. In practice, convergent proofs are  of little help and point to the need for studies that compare different strategies on different sets of data of varying sizes and distributions to evaluate their relative performance in a finite number of iterations or experiments as a function of the dimensionality of the problem, as well as the influence of measurement noise. 
 It has been shown in a number of studies that trade-off strategies, such as EGO and KG perform well in maximizing/minimizing material properties, even for complex systems where the property behavior may include  multiple local maxima/minima.\cite{Fukazawa2019Bayesian, Wang2015NESTED, Yuan2018Accelerated, Balachandran2016adaptive, xue2017informatic}
 However, the performance of these utility functions in the rapid and accurate estimation of  material property curves/surfaces has not been studied.
 Here we propose an active learning loop to 
compare the efficiencies of six utility functions to estimate material property curves in terms of the number of new experiments required for each.

Since selection via maximum variance (\emph{Max-v}.) is one of our utilities, we  introduce the utility, \emph{B.EGO}, designed to minimize/maximize the variability in the function over many bootstrap samples. 
This is in contrast to finding the maximum / minimum  of the function, which is what EGO and KG have so far been applied to. The uncertainties are given by the Kriging model and used in evaluating \emph{Max-v}, \emph{B.EGO}, EGO, KG, random exploration using a uniform distribution, and SKO, Sequential Kriging Optimization,  in recommending the next candidate. The last utility considers the effects of experimental noise on the data. We apply our approach to several  problems with increasing complexity to determine which utilities are robust across all of these. As a common problem in materials science is to predict property curves from limited data, we examine two applications, the fatigue life curve for 304L stainless steel (SS304L) and  estimating the Liquidus line of the Fe-C phase diagram. We show that two or three new experiments or calculations are all that is needed to complete the curve optimization. Since these curves are 1D, we also consider surfaces in the form of the 3D Hartmann 3 function used frequently in optimization tests,  to which we also add  experimental noise to study utility performance for noisy data, and the fitted surface for the intermolecular potential of Ar-SH.   Finally, we apply our tests to a data set of experimental measurements for the Curie temperature of BaTiO$_3$ ferroelectrics, which is modeled in 4D.

\textcolor{black}{Our principal conclusion is that for a range of materials data and problems with varying complexities, directed exploration via maximum variance generally performs better than other utilities. The variability utility \emph{B.EGO} based on bootstrap samples is also a good performer, following \emph{Max-v}. However, 
for given problems, the trade-off methods that add various degrees of exploitation can perform at least as well, if not better, than maximum variance \emph{Max-v}.
Thus, the choice of the utility function is sensitive to  the distribution of the training and  subsequently acquired data, the model performance, the noise as well as the budget, which determines the number of iterations allowed.
}

 \begin{figure*}[htbp]
	\begin{center}
		\includegraphics[width = 18cm]{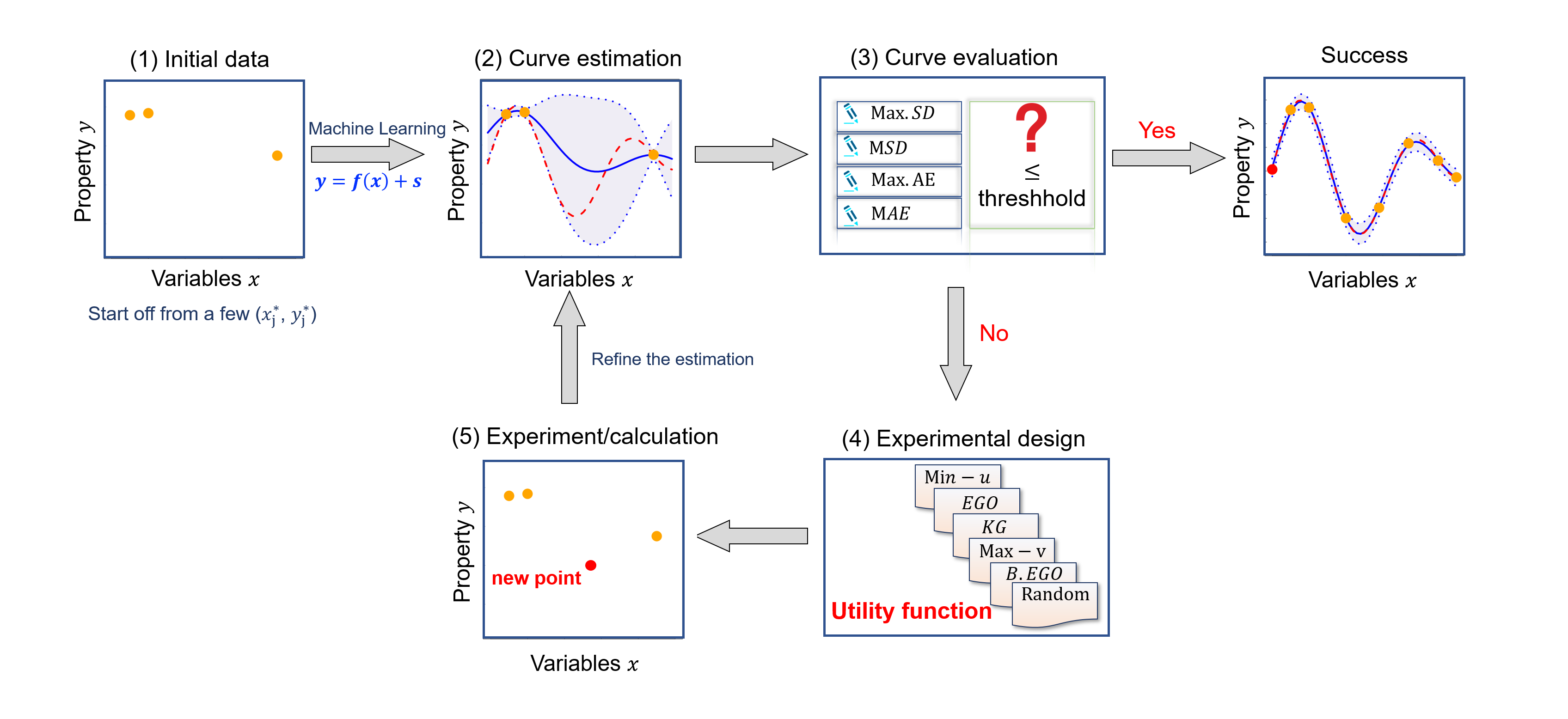}
		\caption{\textcolor{black}{Flow chart of active learning strategy for efficient estimation of material property curves. (1) A small, labeled training data serves as the starting point for the targeted curve. (2) A surrogate model using machine learning methods provides an estimate of the fit with uncertainties. (3) Several criteria are used to check if the estimated fit is adequate and fulfills conditions for success. If not, (4) utility functions are evaluated and ranked to recommend the next data point to use from the pool of possible points for measurement or calculation to determine the label. (5) The recommended experiment/calculation is performed and the new labeled point augments the training data  to obtain a revised estimate of the curve. The loop continues until the criteria for success are met.}
 }
		\label{fig:1}
	\end{center}
\end{figure*}

 \section{Active learning strategy}

\autoref{fig:1} illustrates our active learning loop for material property curve estimation.
We begin with a machine learning model that uses regression to estimate a property curve from the relatively small number of labeled data points available.
The uncertainties associated with this estimation will be large due to limited data.
We employ a Kriging model as the regressor to obtain the predicted value, $\mu$, for each point in the curve as well as the variance of the prediction, $s$, at that point. 
\textcolor{black}{The utility functions are defined in terms of $\mu$ and $s$ and recommend the next unlabeled point for the curve for which the label is evaluated by the user via experiments or calculations. The variance serves as input to Max-v as well as B.EGO. The latter is defined in terms of the bootstrap mean error $\bar s$ and its standard deviation $se(s)$. }
The new point selected then augments the training data so that the regressor can refine and provide an updated value for $\mu$ and $s$.
The loop then continues so that the curve can be improved step by step until an adequate estimate of the
 curve is obtained.
The chief ingredients of our active learning strategy includes the Kriging model, the evaluation criterion of the estimated curve and different utility functions. 

\subsection {Model: Gaussian Process via Kriging }

Machine learning algorithms are efficient at regressing data points to estimate a relationship between a property $y$ and independent variables $x$,  \emph{i.e.}, $y = f(x)$.
They thus serve to build a surrogate model for the estimation and in this work we will use a Kriging model  to perform the regression from the small number of training data points available.
Kriging is a spatial interpolation method that provides a robust estimate of targets and uncertainties associated with unknown points.\cite{CRESSIE1990origins,JOURNEL1977KRIGING, Ginsbourger2010Kriging}  
The interpolated values are modeled by a Gaussian process governed by prior covariances. 
It is customary to consider noisy observations of the targeted property $y$,  where $\tilde{y}_j=y(x_j)+\epsilon_j$ 
and  $\epsilon_j$ is a realization of a random variable
so that $\varepsilon_j$ follows an independent, identically distributed Gaussian distribution $\mathcal{N}(0,\tau_j^2)(1 \leq j \leq n$) with homogeneous noise variance $\tau_j^2$.

The property $y$ is considered a realization of a Gaussian process $Y$ following Kriging. That is,
\begin{equation} 
Y=\bf{m} + \mathnormal Z=\sum \beta \mathrm{f} + \mathnormal Z,
\label{eq:1.1}
\end{equation}
  where $\bf{m}$ is a trend function, $\beta$ is the coefficient and the process $Z$ is assumed Gaussian.

Assuming $p$ training data, $x^*$, with unknown data points, $x$,
 the universal Kriging (UK) equations are 
 given by \cite{Rasmussen06gaussianprocesses, Roustant2012DiceKriging}:
\begin{equation} 
\mu={\bf m}(\mathnormal x)+\mathnormal K(\mathnormal x,\mathnormal x^*)(\mathnormal K+ \Delta)^{-1}(\tilde{\mathnormal y}-\bf{m}(\mathnormal x^*)),
\label{eq:1}
\end{equation}
\begin{equation} 
\begin{aligned}
s^2=s^2_{SK}+(\mathrm{f}(x)^\intercal - \mathnormal K(\mathnormal x,\mathnormal x^*)^\intercal( \mathnormal K+\Delta)^{-1}\mathrm{f}(x^*))^\intercal(\mathrm{f}(x^*)^\intercal \\( \mathnormal K+\Delta)^{-1}\mathrm{f}(x^*))^{-1}(\mathrm{f}(x)^\intercal - \mathnormal K(\mathnormal x,\mathnormal x^*)^\intercal( \mathnormal K+\Delta)^{-1}\mathrm{f}(x^*)),
\label{eq:2}
\end{aligned}
\end{equation}
where $\tilde{y}$=$(\tilde{y}_1,...,\tilde{y}_p)^T$, $K$ is covariance between training data points, $\Delta$ is a diagonal matrix with  diagonal terms  $\tau_1^2,...,\tau_p^2$. 
The simple Kriging (SK) variance, $s^2_{SK}$, is given by
\begin{equation} 
s^2_{SK}=K(x,x)-K(x,x^*)(K+\Delta)^{-1}\mathnormal {K(x^*,x)},
\label{eq:20}
\end{equation}

We use the covariance  kernel $g(h) = exp(-\frac{1}{2}(\frac{h}{\theta})^2)$, where $h$ and $\theta$ are hyper-parameters of the model and set characteristic length-scales associated with the data.
Note that the variance value $s^{2}$ at $x$ depends on the distance from known point $x^*$. 
If $x$ is close to known point $x^*$, it is influenced by $x^*$ and the variance at $\mathnormal x$ will be small.
If  $x$ is separated from known points, the variance at $\mathnormal x$ will be large.

\subsection{Evaluating goodness of fit for the targeted curve}

We determine the quality of the model by tracking the deviation of the regressed curve from the true curve.

In our testing case, as the true curve is known, we can use the mean absolute error (\emph{MAE}) and maximum absolute error (\emph{Max.AE}) defined by
\begin{equation} 
MAE = \frac{1}{n}\sum_{j=1}^{n}(|y_j-\mu_j|),
\end{equation}
\begin{equation} 
Max.AE = max(|y_j-\mu_j|),
\end{equation}
where $n$ is the total number of possible points in the function, $y_j$ are the true values and $\mu_j$ are the estimated values from Kriging model.
The error MAE is the average deviation of the estimate value from the true value whereas Max.AE is the largest error over the range of data points.

As the true curve is usually not known,  we utilize the uncertainty associated with the regressor prediction as an estimate of the model quality. 
We thus use instead the mean standard deviation (\emph{MSD}) and the maximum standard deviation (\emph{Max.SD}) defined as follows
\begin{equation} 
MSD = \frac{1}{n}\sum_{j=1}^{n}(s_j)
\end{equation}
\begin{equation} 
Max.SD = max(s_j),
\end{equation}
where $s_j$ is the standard deviation associated with each prediction ($\mu_j$)  in the curve. 
 We will monitor the evolution of MAE, Max.AE, MSD and Max.SD as we iterate the active learning loop until the accuracy threshold is reached.

\subsection{Utility functions}

Small data sets, characteristics of many materials science problems, typically give rise to large uncertainties in prediction and
therefore additional statistical design criteria need to be invoked. A utility function allows us to choose between experiments or calculations by maximizing an expected utility, where the utilities are defined with respect to information theoretic considerations. By ranking the expected value of the information for possible alternatives for observation or calculation, the utility function  provides the means to prioritize the decision making process  based on the information gained or reduced by observing a potential new data point.  \cite{Casciato2012Optimization,Powell2011Knowledge}
The utilities we compare in this work are defined below, in the noise case we employ \emph{SKO} instead of \emph{EGO}.

{\bf \emph{Min-u}}.
\emph{Min-u} is a greedy choice in which the candidate with the lowest predicted mean value from the model is chosen. That is,
  \begin{equation} 
\nu_{Min-u}= \mu
\end{equation}

{\bf \emph{Max-v}}.
The variability in the predictions can be characterized by the variance at that point obtained from the Kriging covariance \autoref{eq:2}.
Thus, \emph{Max-v} is a risk-averse utility function selecting the next candidate point based on the magnitude of the variance in the property at a given point.
That is, 
 \begin{equation} 
\nu_{max-v}= s
\end{equation}

{\bf \emph{EGO}}.  Efficient Global Optimization (EGO)  balances exploration and exploitation by evaluating the
``Expected Improvement" (\emph{EI}) and choosing the candidate with the largest  (\emph{EI}).
If $\tilde{y}^*_{min}$ is the minimum value in the training data, 
the improvement at a point $x_j$ is $\emph{I}=max(\tilde{y}^*_{min}-Y_j,0)$, where $Y_j$ is distributed normally,  $\mathcal{N}(\mu_j,s_j^2)$.
As the tail of the density function at point $x_j$ extends into $\tilde{y}^*_{min}$,  improvement is then possible.
Different amounts of improvement or distances from $\tilde{y}^*_{min}$ are associated with different density values. 
The EI is obtained by weighting all these improvement values by the associated density values. 
The EI of each potential measurement is the expectation of  \emph{I} at that point given by\cite{Jones1998Efficient}
\begin{equation} 
\begin{aligned}
\nu_{EGO}&=E\left[ max(\tilde{y}^*_{min}-Y_j,0) \right]  = s\mathscr{G}(\frac{\tilde{y}^*_{min}-\mu}{s})\\ &=(\tilde{y}^*_{min}-\mu)\Phi(\frac{\tilde{y}^*_{min}-\mu}{s})+s\phi(\frac{\tilde{y}^*_{min}-\mu}{s})
\end{aligned}
\end{equation}
where $\mathscr{G}(z_0) = z_0\Phi(z_0) +\phi(z_0)$, $z_0 = \frac{\tilde{y}^*_{min}-\mu}{s}$, $s$ is the standard deviation associated with the mean value $\mu$ of the model prediction, $\Phi(\cdot)$ and $\phi(\cdot)$ are the standard normal density and distribution functions, respectively.
If the measurements are noise-free,  $\nu_{EGO}$ is zero at the sampled points (points that are already measured) and is positive in elsewhere. 
In our discretized version of the problem here, EGO simply evaluates \emph{EI} at each unexplored point and recommends a point with the largest \emph{EI} to be measured next.

\textcolor{black}{{\bf \emph{SKO}}.
In the noise case, the current best estimate $\tilde{y}^*_{min}$ also suffers from noise, and the actual minimum is indeed unknown.
 We therefore utilize an extension of \emph{EGO}, Sequential Kriging Optimization (\emph{SKO}) \cite{Huang2006Global}, in which $\tilde{y}^*_{min}$ in \emph{EGO} is modified through the model predictions $\mu^{**}$.  A prefactor $1-\frac{\epsilon}{\sqrt{\epsilon^2+s^2}}$ is introduced to enhance the exploration. 
}
\begin{equation} 
\begin{aligned}
\textcolor{black}{\nu_{SKO}=(1-\frac{\epsilon}{\sqrt{\epsilon^2+s^2}}) s \mathscr{G}(\frac{\mu^{**}-\mu}{s})}
\end{aligned}
\end{equation}
\textcolor{black}{where $\mu^{**} = \mu (argmin \left[\mu+\lambda s \right])$, $\lambda$ is the ``risk avoidance'' parameter.}

{\bf \emph{KG.}} 
Knowledge Gradient(KG) aims at maximizing the current reward\cite{Frazier2008KNOWLEDGE,Frazier2009Correlated}.
KG can be calculated using
\begin{equation} 
\begin{aligned}
\nu_{KG} = S\mathscr{G}(-|\frac{\mu'-\mu}{S}|)
\end{aligned}
\end{equation}
where $\mu'$ is the minimum of the predicted values in the virtual unexplored space ($\mu_j$) without the next recommended sample ($\mu_i$), {\it i.e.,} $\mu' = min\mu_j$, for $j \ne i$;
\textcolor{black}{If the measurements are noise-free, $S$ is the standard deviation $s$ provided by  the Kriging model directly.
KG is also available for the noise case} and the modified standard deviation $S$ is given by 
${S=s^2/\sqrt{\tau^2+s^2}}$, where $s^2$ is the variance of the prediction and $\tau^2$ is the variance associated with noise. \cite{Frazier2008KNOWLEDGE,Frazier2009Correlated, Powell2011Knowledge}

\textcolor{black}{{\bf \emph{B.EGO}}.}
\textcolor{black}{In order to find the point with the maximum variability in the property using EGO or KG, we introduce a new utility that complements \emph{Max-v}.  Whereas the  variability in \emph{Max-v} is directly obtained from the Kriging model for a given data set,  the  {\it variability} in \emph{B.EGO} is obtained by considering many bootstrap samples.}
\textcolor{black}{Let $s^*_{max}$ be the largest so-far value of the standard error of the bootstrap uncertainties in the training data, so that 
 $\emph{I}=max(ERROR_j-s^*_{max},0)$, where $ERROR_j$ is distributed normally,  $\mathcal{N}(\bar s_j,{se(s)}_j^2)$.
The mean  $\bar s$ is given by  $\frac{\sum_{b=1}^B s_b}{B}$, where B are the bootstrap replicates or samples, and 
$se(s)$ is the standard error of the bootstrap uncertainties corresponding to $\bar s$. We use a value of B of  50.  
Then the expected improvement, EI of each potential measurement is the expectation of  \emph{I} at that point given by\cite{Jones1998Efficient}}
\begin{equation} 
\begin{aligned}
\nu_{B.EGO}&=E\left[ max(ERROR_j-s^*_{max},0) \right]\\  &= se(s)\mathscr{G}(\frac{\bar s-s^*_{max}}{se(s)})\\ &=( \bar s-s^*_{max})\Phi(\frac{\bar s-s^*_{max}}{se(s)})+se(s)\phi(\frac{\bar s-s^*_{max}}{s})
\end{aligned}
\end{equation}
\textcolor{black}{where $\mathscr{G}(z_0) = \int^{z_0}_{-\infty}\Phi(z)dz = z_0\Phi(z_0) +\phi(z_0)$, $z_0 = \frac{\bar s-s^*_{max}}{se(s)}$. 
\emph{B.EGO} is designed  to aim at searching for the point with the maximum Expected Improvement, however, the improvement is not for the current minimum function value but for the current maximum standard deviation of all labeled observations. }

{\bf \emph{Random}}. This involves a random choice of  the unmeasured candidate, such that if there are a total of N choices,  $x_i$ is chosen with probability 1/N. 

\section{results}

 We present results for three classes of problems with varying complexity.  \textcolor{black}{We consider 1D and multi-dimensional cases with a finite number of measurements of relevance in materials science. 
 These  include Case I: a) the fatigue life cycle curve for  304L stainless steel (SS304L) and b) the Liquidus line in the Fe-C phase diagram, 
Case II: a) a standard optimization test function, the  Hartmann 3 function in 3D, and b) the fitted intermolecular potential surface for Ar-SH, 
 and Case III: Measurements of the Curie temperature for ferroelectric samples with 4 variables or features. 
In Case II a) we also vary the data set size in the presence of ``experimental" noise.}
In each case a small subset $\tilde{y}^*_j$ and $x^*_j$ of the data set  is randomly chosen as the initial training data
and the remaining data $\tilde{y}_j$ and $x_j$ comprise the unexplored search space.
We implemented the feedback loop of \autoref{fig:1}, monitoring the departures from the true result using uncertainties given by \emph{MAE}, \emph{Max.AE}, \emph{MSD} and \emph{Max.SD} defined previously.
We employed Kriging to perform regression  and 
once the new measurement $x_i$ is selected by the utility function, the new observed value $\tilde{y}_i$ augments the training data and the loop repeats itself, refining successive estimates.
We monitor the number of iterations, ($\mathsf{N}$), of the loop, {\it i.e.}, number of new measurements made with a given utility that minimize ($\mathsf{N}$).
To garner adequate statistics, the design process was repeated 100 times with different initial training data randomly selected from all the discretized points in the data set.

\subsection{Case study I: 1D materials cases}

\noindent {\bf a) Fatigue Life Curve for 304L stainless steel.}

Fatigue properties of materials are often described using the fatigue curve, which
describes the relationship between cyclic stress and number of cycles to failure. It is critical in assessing material failure and
obtaining it experimentally requires a series of tests to find the ultimate stress for a given number of loading cycles, quite time and cost consuming.

\begin{figure*}[htbp]
	\begin{center}
		\includegraphics[width = 12cm]{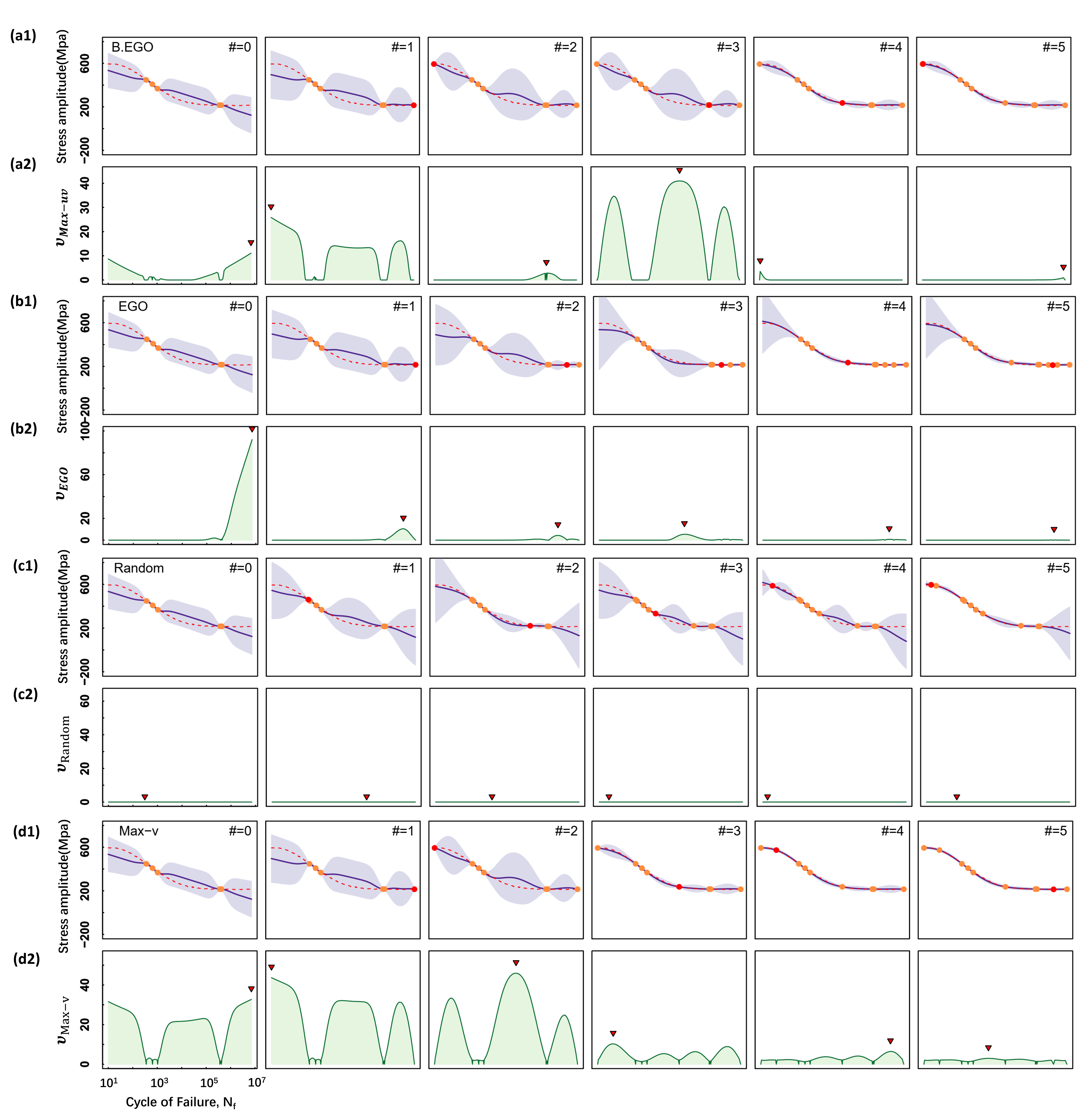}
		\caption{
		\textcolor{black}{Optimization process for the Fatigue Life Curve for SS304L steel comparing the behavior of different utility functions. 
		The solid blue line is the Kriging mean estimate and to visualize the uncertainties, the shaded region denotes 10 times the standard deviation at each input value. The  red arrow denotes the next infill point (maximum) in the utility functions. The optimizations are initialized with the same 5 points, but quickly follow different sampling trajectories. }}
		\label{fig:9}
	\end{center}
\end{figure*}

We choose a monotonic fatigue life curve for 304L stainless steel from the simulation work of F.Mozafari et al.\cite{Mozafari2019rate} as a typical example to validate our design loop in \autoref{fig:1}.
The dotted red line in \autoref{fig:9} shows this curve.
We consider the number of cycles to failure, N$_f$, as the independent variable and the stress amplitude $\sigma_a$ as the output or property and discretize the curve into 201 data points. 
We randomly choose 5 data points as the initial training data with known $x^*_j$ and $\tilde{y}^*_j$  to employ in the loop of \autoref{fig:1} to optimize the curve.
The next measurement is recommended from the remaining 196 data points using different utility functions.  

Typical examples of the optimization process from one initial dataset are shown in \autoref{fig:9}.
The panels (\# = 0) of \autoref{fig:9}  show the Kriging interpolation estimate of the curve from the 5 initial data points for the different utility functions. 
The predicted values $\mu_j$ of all the unexplored/explored $x_j$ are shown by the solid blue line with the shadows about the solid blue line tracking the uncertainties $s_j$ associated with  $\mu_j$.
The green curves in  \autoref{fig:9} (for example a2 corresponding to the curve a1) show the behavior of the utility function ($\nu$)  for each point on the curve ($x_j$).
The $x_i$ with the largest value of the utility function $\nu_i$ (indicated by the red arrow) is recommended for the next measurement.
The Kriging model is then refined based on 6 data points, and the updated curve is shown in  panel (\# = 1) of \autoref{fig:9} .
The  process is repeated 5 times for the different utility functions and the changes due to new selections of points are quite apparent from panel to panel in \autoref{fig:9}.
For each utility function in  \autoref{fig:9}, the optimization begins with the same 5 points but is followed by different sampling trajectories. \autoref{fig:9} shows a typical example of the optimization process for  \emph{Max-v}, Random compared to EGO and \emph{B.EGO}.
\textcolor{black}{The function \emph{Max-v} converges to the true objective function in only three new measurements, outperforming the other functions which need more measurements.}

\begin{figure}[htbp]
	\begin{center}
		\includegraphics[width = \columnwidth]{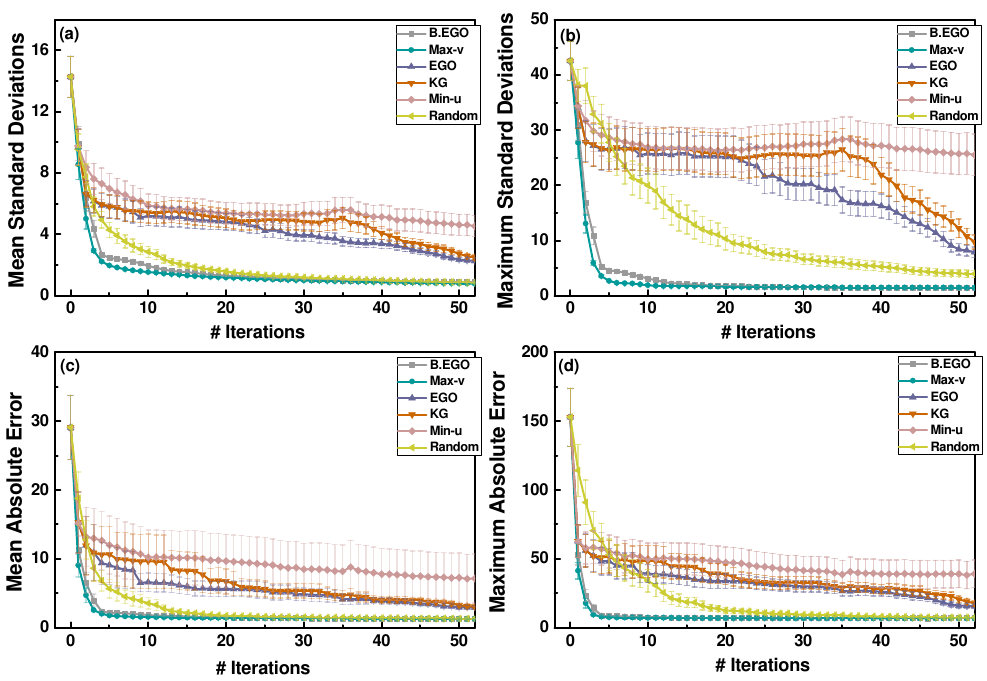}
		\caption{
		\textcolor{black}{Comparison of the performance of utility functions in optimizing the Fatigue Life Curve for SS304L steel. The initial data size contains $n$ = 5 training points and  the mean values and  error bars showing the 95\% confidence levels of the points are evaluated over 100 trials. Shown is the behavior of (a) mean standard deviation,  (b) maximum standard deviation,  (c) mean absolute error, and (d) maximum absolute error.}}
		\label{fig:8}
	\end{center}
\end{figure}

For a more robust comparison, we used an initial random training data set of $n$ = 5 training points and tracked the values of \emph{MAE}, \emph{Max.AE}, \emph{MSD} and \emph{Max.SD} as a function of number of new measurements for the different policies. By repeating over a 100 trials, \autoref{fig:8} (a) - (d) shows the  average values and 95\% confidence intervals for \emph{MAE}, \emph{Max.AE}, \emph{MSD} and \emph{Max.SD} for the different utility functions.
\textcolor{black}{
Both  \emph{Max-v} and our new function \emph{B.EGO} perform very well, converging in relatively few iterations followed by   \emph{Random} which also converges but with more iterations.
The trade-off methods \emph{EGO} and \emph{KG} decrease the error quickly in the first three iterations, but then relax slowly but nevertheless also converge.
The greedy, pure exploitation \emph{Min-u} shows very little relaxation after a few iterations. }

\noindent { \bf b) Liquidus line in the Iron-carbon phase diagram.}

\begin{figure*}[htbp]
	\begin{center}
		\includegraphics[width = 12cm]{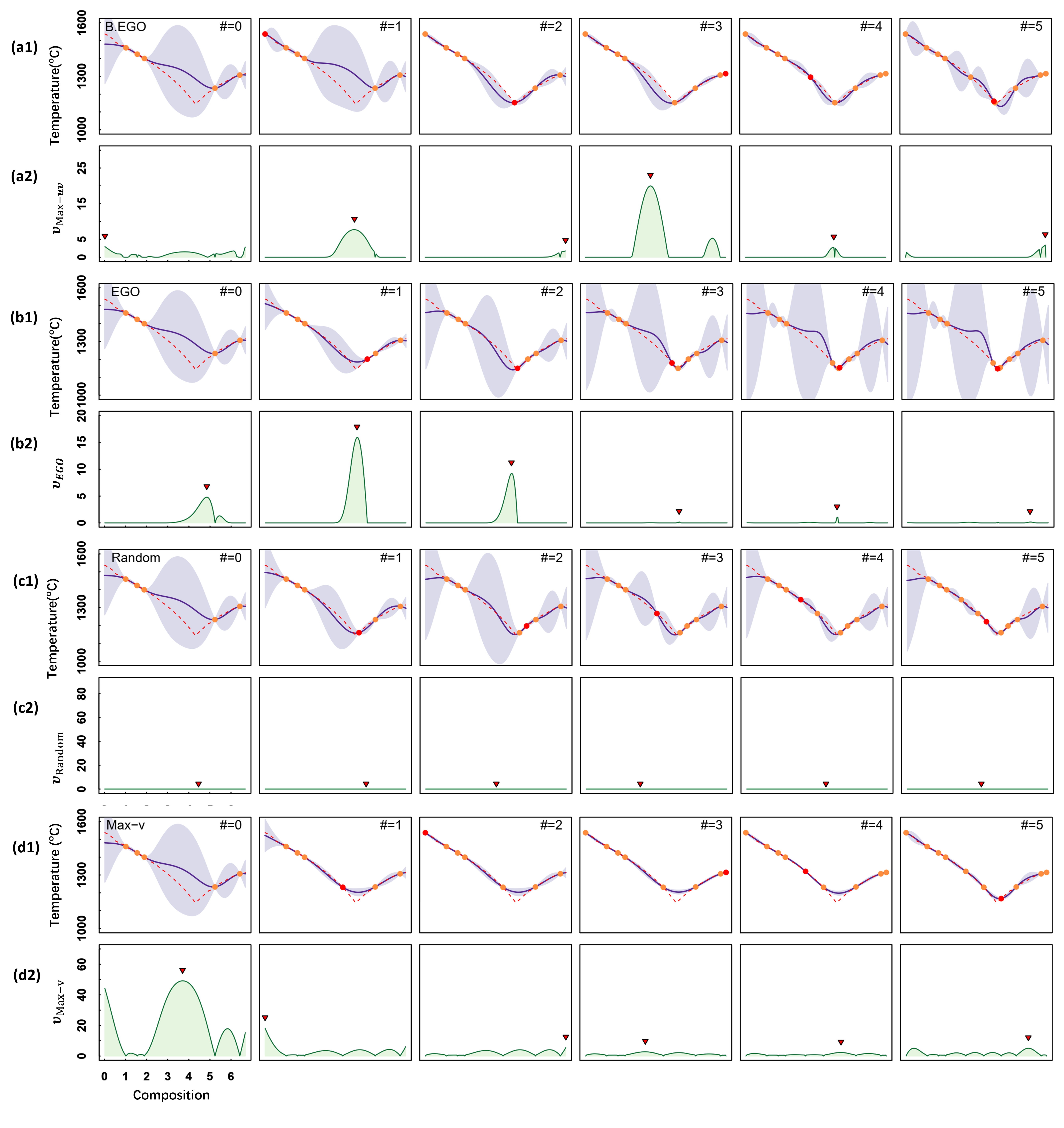}
		\caption{
		Optimization process for the Liquidus line (red dotted) of the Fe-C phase diagram comparing the effects of different utility functions. 
		The solid blue line is the Kriging mean estimate and to visualize the uncertainties, the shaded region denotes 10 times the standard deviation at each input value. The  red arrow denotes the next infill point (maximum) in the utility functions. The optimizations are initialized with the same 5 points, but quickly follow different sampling trajectories. }
		\label{fig:7}
	\end{center}
\end{figure*}

The Iron-carbon (Fe-C) phase diagram displays the phases, compositions and transformations in iron-carbon alloys as a result of heating and cooling, 
and therefore serves as the basis for composition design and optimizing heat treatment of steels.
The liquidus line is the phase boundary in the phase diagram limiting the bottom of the liquid field, and the 
dotted red line in \autoref{fig:7} shows the Liquidus line exhibiting a eutectic point at C composition of 4.3\% between $\gamma$ and Fe$_3$C.
The  curve is irregular and the challenge is to obtain it with as few measurements as possible.
 We discretize the Liquidus line into 118 data points, {\it i.e.}, 118 composition-temperature data points and
 randomly choose 5 initial points.
The true Liquidus line of the Fe-C phase diagram is shown by the dotted red line.  
The estimated curves initially deviate significantly from the true curve, which gives rise to large values of \emph{MAE}, \emph{Max.AE}, \emph{MSD} and \emph{Max.SD}.
In \autoref{fig:7}, the function \emph{Max-v} only requires two new measurements to match the true function and  outperforms all the the other functions which need more measurements.
 The function \emph{B.EGO} also does well in the optimization as it  works directly on the prediction errors. 
Both  \emph{EGO} and \emph{Random} predict a curve  close to the true function in the second iteration, but then get worse as the iteration number increases.

\begin{figure}[htbp]
	\begin{center}
		\includegraphics[width = \columnwidth]{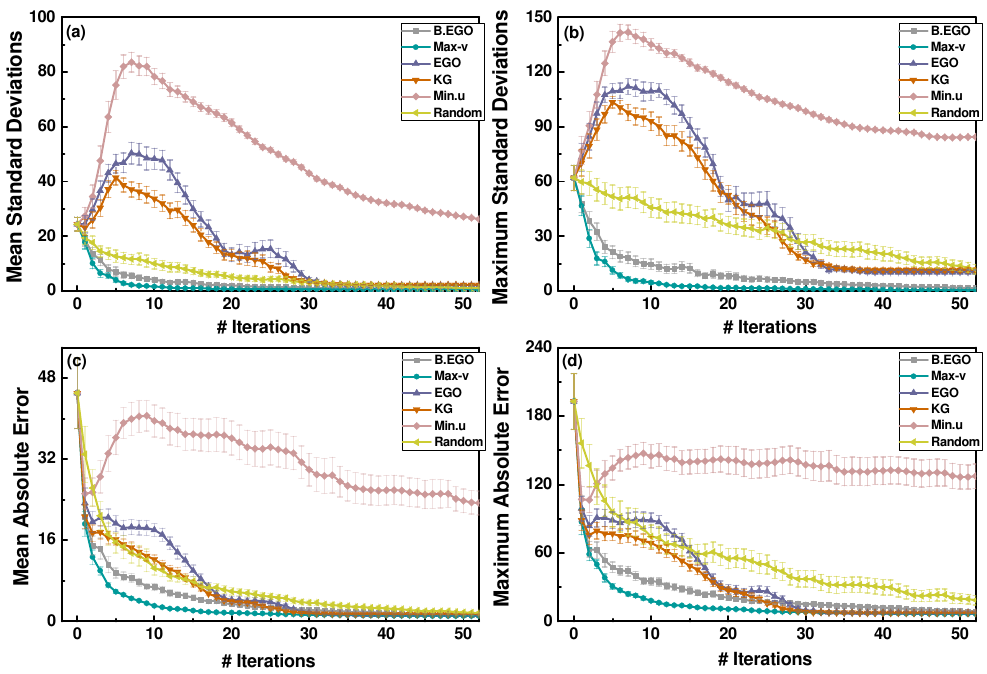}
	 in 	\caption{
		\textcolor{black}{Comparison of the performance of utility functions in optimizing the Liquidus line of the Fe-C phase diagram. The initial data size contains $n$ = 5 training points and  the mean values and  error bars showing the 95\% confidence intervals of the points are evaluated over 100 trials. Shown is the behavior of (a) mean standard deviation,  (b) maximum standard deviation,  (c) mean absolute error, and (d) maximum absolute error.}}
		\label{fig:6}
	\end{center}
\end{figure}

\textcolor{black}
{
By repeating over a 100 trials, \autoref{fig:6} (a - d) shows the  average values and 95\% confidence intervals for \emph{MAE}, \emph{Max.AE}, \emph{MSD} and \emph{Max.SD} for the different utilities.
\autoref{fig:6} essentially bears out our previous findings for the fatigue curve seen in \autoref{fig:8}, 
and the general features are very similar to those discussed previously for the fatigue curve.
The uncertainty based \emph{Max-v} and \emph{B. EGO} perform well and converge readily compared to Random and the trade-off methods, all of which do converge although require more iterations. \emph{Max-v} relaxes more quickly than \emph{B. EGO}
if compared to fatigue, however, other than pure exploitation \emph{Min-u}, all the exploratory utilities (including random) converge in the 1D materials data sets.
}

 \textcolor{black}{\subsection{Case study II: Higher dimensional surfaces}}
  
 \textcolor{black}{\noindent {\bf a) Hartmann 3 function}}

\label{case4}

\begin{figure*}[htbp]
	\begin{center}
		\includegraphics[width = 14cm]{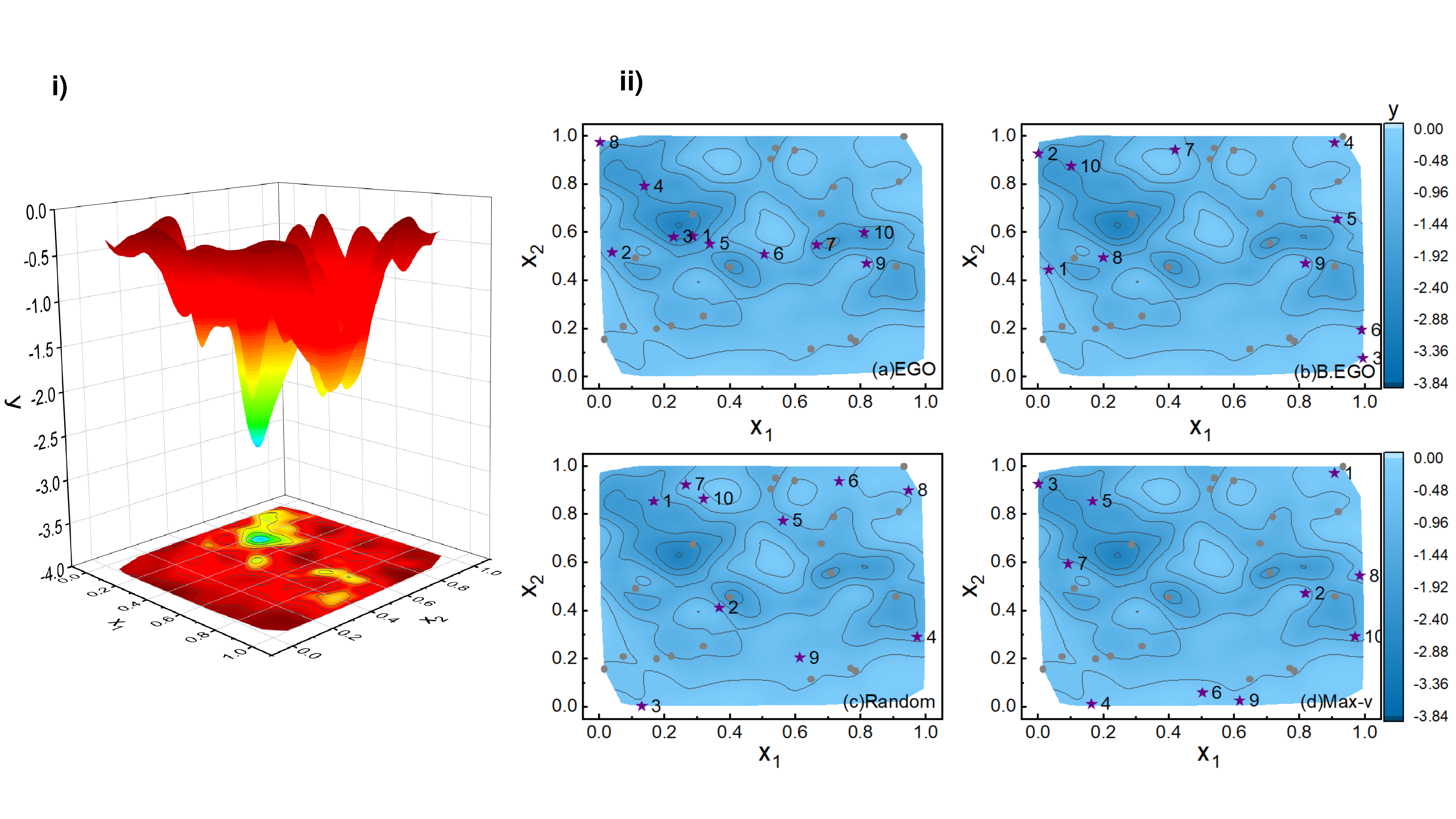}
		\caption{
		\textcolor{black}{i) The surface and contours for the Hartmann3 function constructed using 400 points with independent variables $x_{1}$, $x_{2}$, $x_{3}= YYZZ$. A global minimum and  four local minima are shown.  As the figures are constructed using Latin hypercube sampling,  the minima deviate slightly from the actual minimum of the function. 
		ii) Optimization process comparing the behavior of different utility functions.  Sequences of visited points after 10 iterations for different utility functions are shown by stars with an iteration number. The gray filled circles represent the training data.}
  }
		\label{fig:16}
	\end{center}
\end{figure*}

\textcolor{black}{
We utilize a well-known optimization test function, the Hartmann 3 function, to generate data for a 3-Dimensional mathematical case with multiple local minima and 1 global minimum.
The function is defined by:}
\begin{equation} 
\begin{aligned}
y &= - \sum_{n=1}^4 \alpha_n exp(-\sum_{m=1}^3 A_{nm}(x_m-P_{nm})^2),\\
where \; \alpha &= (1.0, 1.2, 3.0, 3.2)^{\top}\\
A&=\left( \begin{array}{ccc} 3.0 & 10 & 30 \\
0.1 & 10 & 35 \\
3.0 & 10 & 30 \\
0.1 &10 &35 \end{array} \right)\\
P&=10^{-4}\left( \begin{array}{ccc} 3689 & 1170 & 2673 \\
4699 & 4387 & 7470 \\
1091 & 8732 & 5547 \\
381 &5743 &8828 \end{array} \right)
\end{aligned}
\end{equation}

\textcolor{black}{
The whole space is discretized into 400 points ($x_{mj}, m=1,2,3; j=1,2...400$) using Latin Hypercube Sampling.
Their corresponding $y_j$ is the value evaluated by the function.
The surface as well as its projected contours on the two of these variables planes is shown in \autoref{fig:16} i) for the 400 points.
We randomly select 80 data points (20\% of the total search space) as the initial training data, with known $x_{mj}^*$ and $y_{j}^*$, shown by the gray points in \autoref{fig:16} ii).
We ran 10 steps of different utility functions and show the optimization process in \autoref{fig:16} ii).
The stars with the numbers refer to the sequence obtained.
From the distribution of the new 10 points, we can see that the points chosen by \emph{Max-v} and \emph{B.EGO} are widely distributed on the whole surface and initially even points on the edge of the contours are sampled.
With \emph{EGO} very few points are distributed away from the local or the global minimum.}

\begin{figure}[htbp]
	\begin{center}
		\includegraphics[width = \columnwidth]{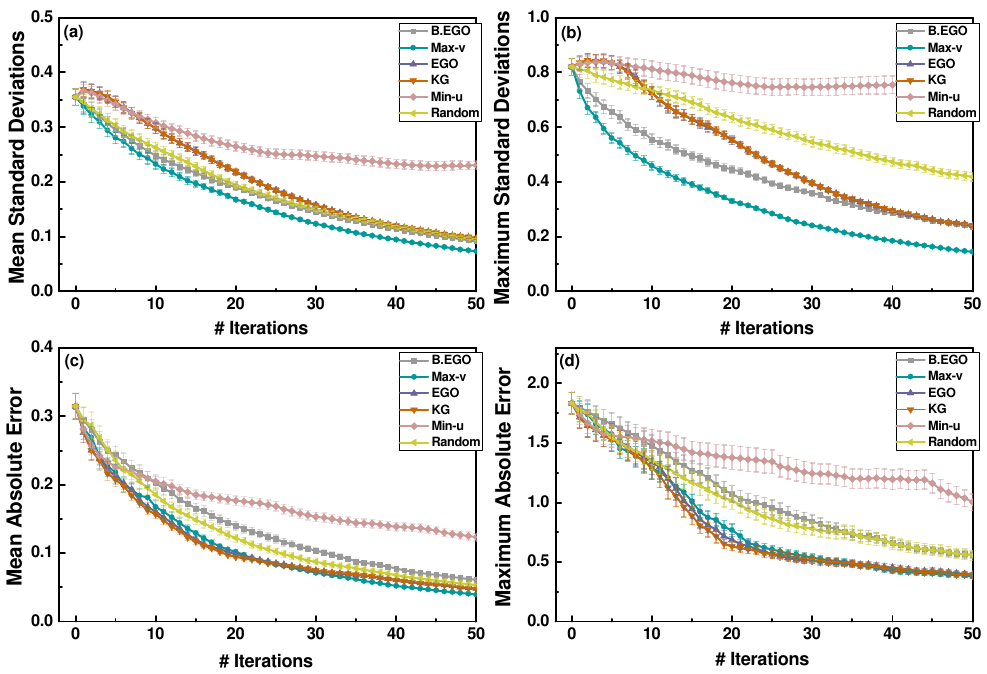}
		\caption{
		\textcolor{black}{Comparison of the performance of utility functions in optimizing the the targeted objective function. The initial data size contains $n$ = 80 training points and  the mean values and  error bars showing 95\% confidence intervals of the points are evaluated over 50 trials. Shown is the behavior of (a) mean standard deviation,  (b) maximum standard deviation,  (c) mean absolute error, and (d) maximum absolute error.}}
		\label{fig:13}
	\end{center}
\end{figure}

\textcolor{black}{To compare the efficiency of  the utility function introduced above in the optimization process, we used an initial randomly training data set of $n$ = 80 training points and tracked the values of \emph{MAE}, \emph{Max.AE}, \emph{MSD} and \emph{Max.SD} as a function of number of new measurements for the different policies.
\autoref{fig:13} (a) - (d) shows the  average values and standard deviation for \emph{MAE}, \emph{Max.AE}, \emph{MSD} and \emph{Max.SD} for 100 trials.
 \emph{Max-v} and the trade-off policies perform better than the rest, including  \emph{B.EGO} and Random. 
This example also suggests that the actual variance lends itself better to exploration of the space than the variability across bootstrap samples.
The performance of \emph{EGO}, \emph{KG} is slightly better than \emph{Max-v} for 30 or less iterations, which can be explained through the optimization sequence shown in \autoref{fig:16} ii).
We observe that the training data are further away from the global minimum in \autoref{fig:16} ii),  suggesting that the points near the minimum will not be well predicted.
\emph{EGO} samples some points near the global minimum in the first few iterations (shown by dark purple stars), whereas  \emph{Max-v}, \emph{B.EGO} and \emph{Random} do not sample points in this area.
Thus, it is not surprising that \emph{EGO} does better initially.
}

\begin{figure}[htbp]
	\begin{center}
		\includegraphics[width = \columnwidth]{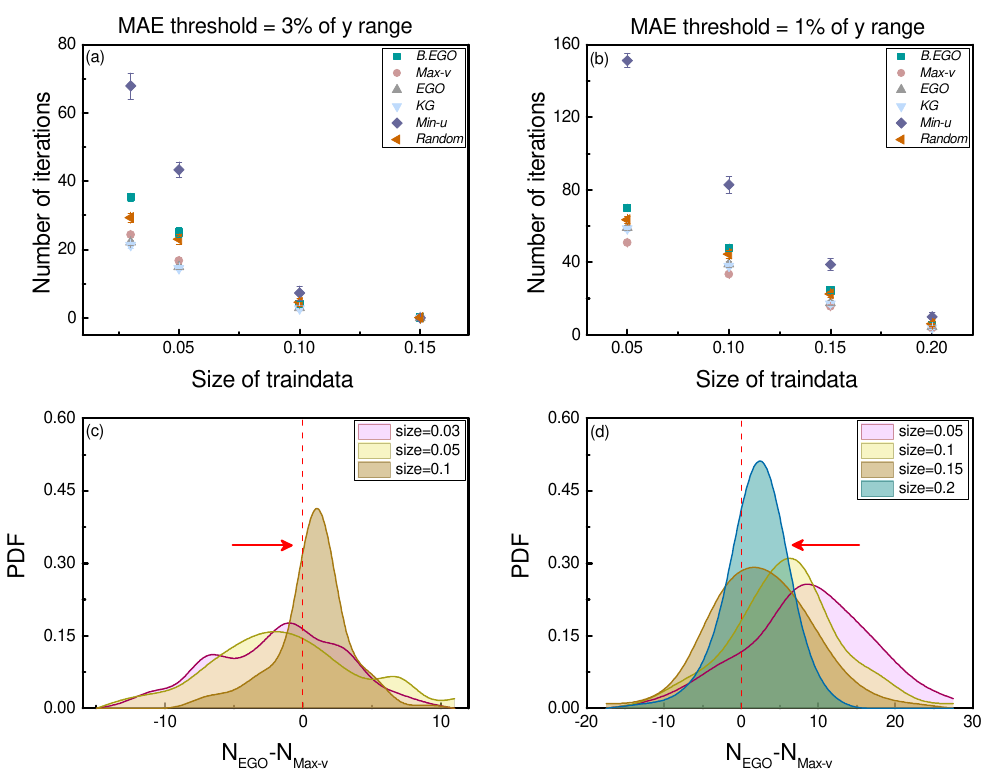}
		\caption{\textcolor{black}{The number of new measurements needed to achieve a given accuracy (1\% and 3\% deviation in maximum and minimum values in \emph{MAE}) for the curve as a function of the training data size(\autoref{fig:5}(a,b)). Each optimization process is repeated  100 times. The points with the error bar are the mean values associated with 95\% confidence levels over 100 trials.
		 The probability density of the difference in the number of iterations using \emph{EGO} and \emph{Max-v}. Plotted along the x axis is $N_{EGO} -  N_{Max-v}$.
		} }
		\label{fig:5}
	\end{center}
\end{figure}

\textcolor{black}{To study the number of iterations required for different policies or utility functions to match a targeted objective function or curve, we set a threshold on \emph{MAE} to stop our iteration loop. 
The threshold is set to ($y_{max}$ - $y_{min}$) $\times$  3\% and 1\% respectively, to show how these utility functions perform to meet different demands of accuracy (shown in \autoref{fig:5}(a,c) and \autoref{fig:5}(b,d)).
The initial training data with sizes from 3\% to 15\% times the number of total data for the first threshold, and 5\% to 20\% times the number of total data for the second threshold was selected randomly.
A total of 200 iterations is set to stop the optimization loop.
If after 200 iterations the loop does not reach the threshold,  the number of new measurements needed is counted as 200.
\autoref{fig:5}(a,b) shows the number of iterations required to meet the threshold as a function of initial training data size. 
Each point with the error bar represents the average value associated with 95\% confidence level over 100  random trials.
As the size of training data increases,  all the utility functions perform better.
 Pure exploitation \emph{Min-u} performs much worse than the others, almost 2 or 3 times slower, followed by \emph{B.EGO} and \emph{Random}.
 However, we notice the differences between the trade-off methods and \emph{Max-v} when the MAE threshold differs.
 If the MAE threshold equals 3\% of y range, the trade-off methods \emph{EGO} and \emph{KG} are a little better than \emph{Max-v}.
  If the MAE threshold equals 1\% of y range, a higher requirement on the accuracy, then \emph{Max-v} is the best choice.
 \autoref{fig:5}(c,d) show the probability density difference ($\Delta$) in iterations needed to reach the threshold using \emph{EGO} and \emph{Max-v} in 100 trials.
 The peak moves from  negative to zero and becomes narrower as the size of training data increases in \autoref{fig:5}(c), indicating that the advantages of \emph{EGO} decrease with the increasing size until they finally perform similarly.
 For MAE threshold equals to 1\% of y range, the opposite of (\autoref{fig:5}(d)) applies.
 }

\begin{figure}[htbp]
	\begin{center}
		\includegraphics[width = \columnwidth]{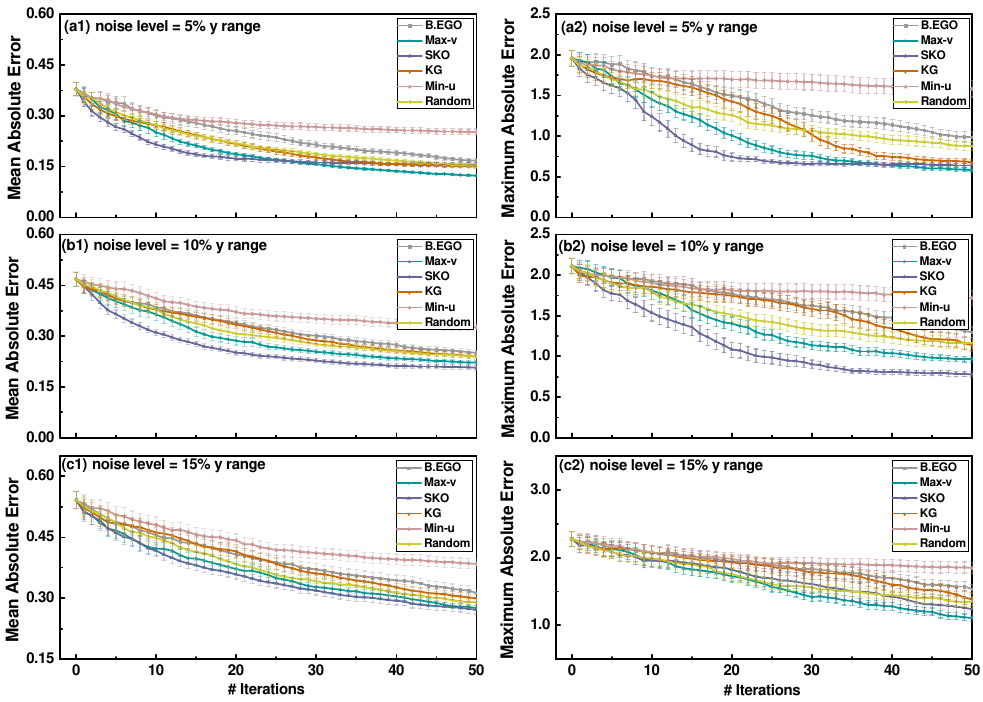}
		\caption{
		\textcolor{black}{Comparison of the performance of utility functions in optimizing the the targeted objective function subject to noise levels of 5\%, 10\% and 15\% corresponding to \autoref{fig:14}(a1,a2), (b1,b2) and (c1,c2). The initial data size contains $n$ = 80 training points and  the mean values and  error bars showing the 95\% confidence intervals of the points are evaluated over 100 trials. Shown is the behavior of (a) mean standard deviation,  (b) maximum standard deviation,  (c) mean absolute error, and (d) maximum absolute error.}}
		\label{fig:14}
	\end{center}
\end{figure}

\textcolor{black}{\noindent {\bf Effects of noise.}}
\textcolor{black}{
Using the function above, we introduce random errors to generate noisy data to simulate noisy measurements in experiments.
We assume noise $\epsilon _j$ follows a normal distribution $\mathcal{N}(0,\tau^2)$, where $\tau$ is set to 5\%, 10\% and 15\% of y range, respectively. 
The observation values then can be calculated via $\tilde{y}_j=y(x_j)+\epsilon_j$.
A second measurement of the same candidate $j$, which has already been measured, is allowed.
The results for the different utility functions after 100 trials are presented in \autoref{fig:14}.
Each row  corresponds to one level of noise, the noise level increases from top to bottom.
Compared to \autoref{fig:13}, if the \# Iterations equals to zero, {\it MAE} and {\it Max.AE} increase.
The increase in  these values in the beginning indicates that noise makes the prediction of the model deviate much more from the real curve.
That is, the prediction suffers from both model uncertainties and measurement noise.
\emph{SKO}, the modified version of \emph{EGO} with noise, performs very well, especially at noise levels of 5\% and 10\%.  {\it Max-v} does surprisingly well and does  converge to \emph{SKO} with more iterations. 
}

\textcolor{black}{\noindent {\bf b) Intermolecular potential energy surface for Ar-SH}}

\begin{figure*}[htbp]
	\begin{center}
		\includegraphics[width=14cm]{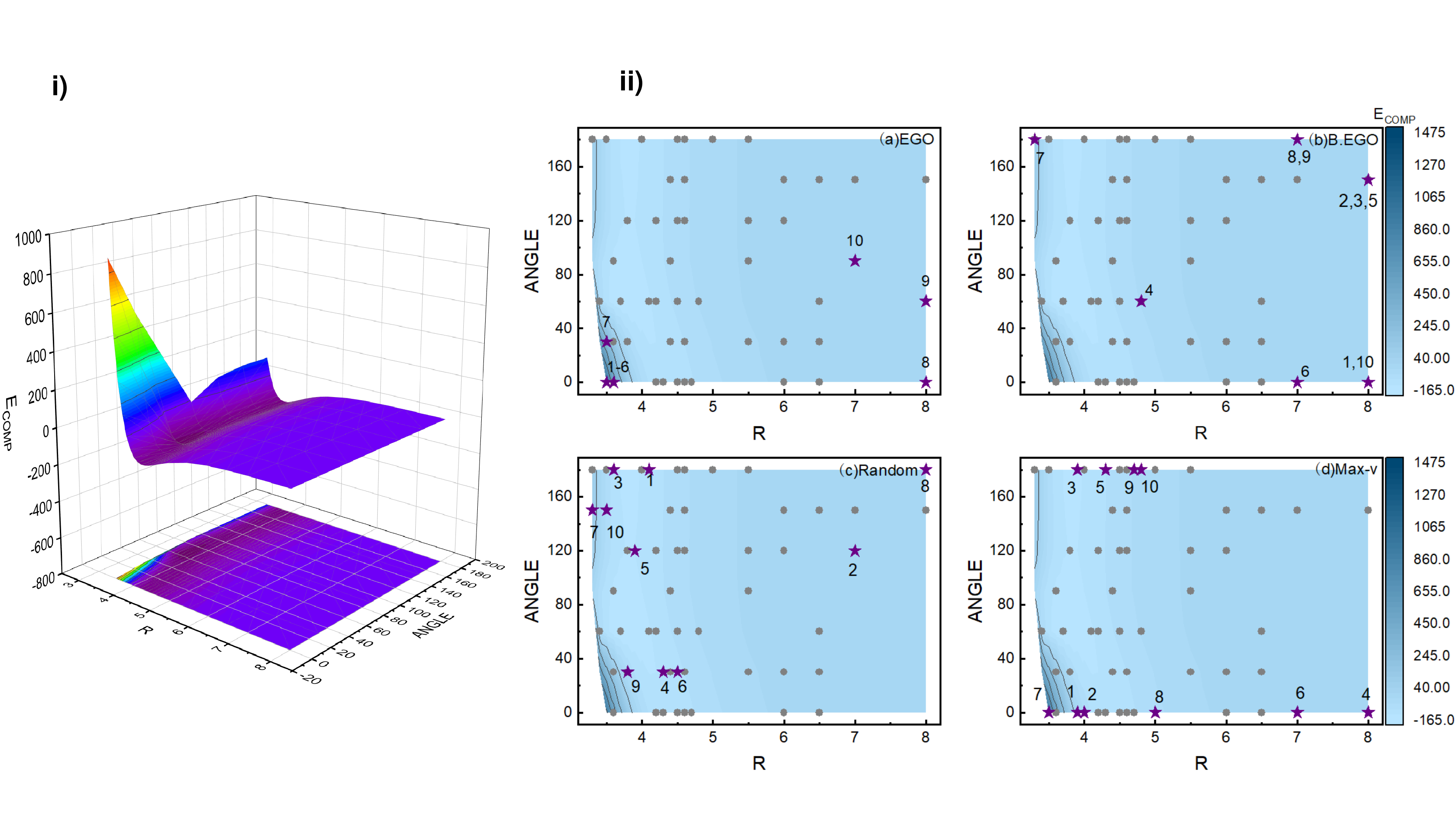}
		\caption{\textcolor{black}{i) The surface and contours of the Ar-SH potential energy surface with independent variables including the distance between Ar and the center of mass of SH in Angstrom(R) and the angle theta(ANGLE). ii) Optimization process comparing the behavior of different utility functions.  Sequences of visited points after 10 iterations for different utility functions  are shown by stars with an iteration number. The gray filled circles represent the training data.}}
		\label{fig:17}
	\end{center}
\end{figure*}

\textcolor{black}{The 3D intermolecular potential energy surface for Ar-SH has been determined by a combination of spectroscopic measurements and solutions to the Schr\"{o}dinger equation \cite{Sumiyoshi2005Spectroscopy}. The fitted surface and potential well reproduces all the known experimental data and we utilize this example to test the utility functions.
Our database includes in total 1050 points with the calculated potential energy in $cm^{-1}$ and 3 variables, namely, the distance between Ar and the center of mass of SH in Angstrom, the angle theta, and the SH bond length in Angstroms.
The surface as well as its projected contours on the two of these variables planes are shown in \autoref{fig:17} i) for the 1050 points.
We randomly selected 52 data points (5\% of the total search space) as the initial training data, shown by the gray solid filled points in \autoref{fig:17} ii).
We ran 10 iterations for each utility function and the sequence of optimizations is shown in \autoref{fig:17} ii).
The stars with the numbers refer to the sequence obtained.
The distribution of the newly acquired 10 starred points by the utility functions is very similar to that for the Hartmann 3 function.
That is, those chosen by \emph{Max-v} and \emph{B.EGO} are widely separated, whereas for \emph{EGO} few points are dispersed away from the local or  global maxima.}

\begin{figure}[htbp]
	\begin{center}
		\includegraphics[width= \columnwidth]{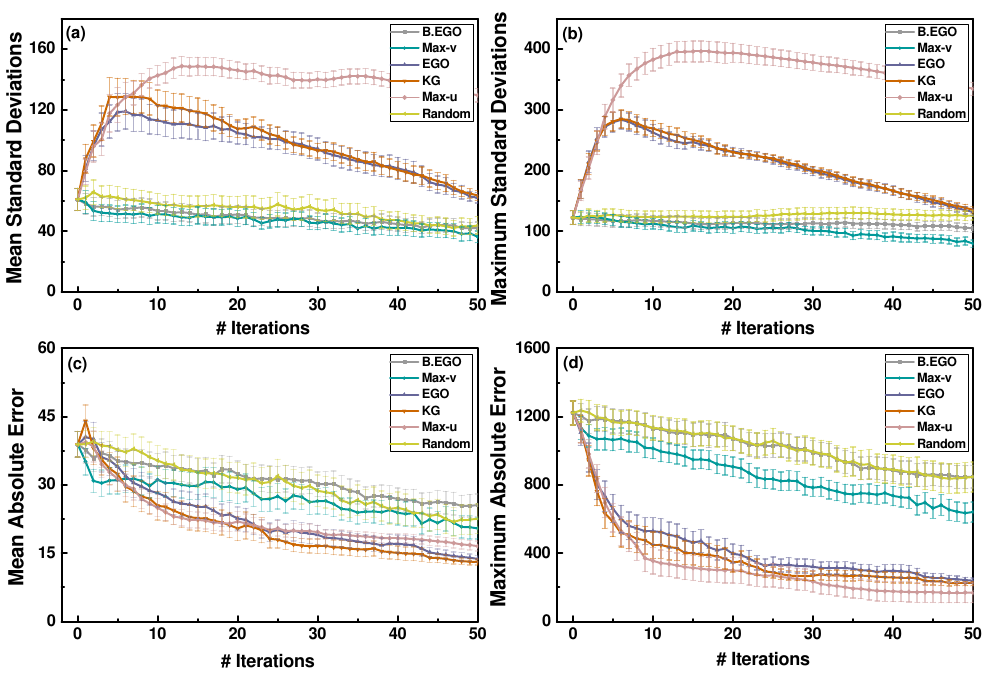}
		\caption{\textcolor{black}{Comparison of the performance of utility functions in optimizing the the targeted objective function. The initial data size contains $n$ = 52 training points and  the mean values and  error bars showing 95\% confidence intervals of the points are evaluated over 100 trials. Shown is the behavior of (a) mean standard deviation,  (b) maximum standard deviation,  (c) mean absolute error, and (d) maximum absolute error.}}
		\label{fig:18}
	\end{center}
\end{figure}

\textcolor{black}{To compare the efficiency of  the utility function introduced  in the optimization process, we used an initial random training data set of $n$ = 52 training points and monitored the values of \emph{MAE}, \emph{Max.AE}, \emph{MSD} and \emph{Max.SD} as a function of number of new measurements for the different policies.
\autoref{fig:18} (a) - (d) shows the  average values and standard deviations for \emph{MAE}, \emph{Max.AE}, \emph{MSD} and \emph{Max.SD} for 100 trials.
The performance of \emph{EGO}, \emph{KG} and \emph{Max-u} is considerably better than \emph{Max-v} in the first 50 iterations shown, and the optimization sequence in \autoref{fig:17} ii) shows the evolution.
As the training data are further away from the global maximum in \autoref{fig:17} ii),  we expect the predictions to have large uncertainties.
\emph{EGO} samples  points near the global maximum in the first few iterations (shown by dark purple stars), whereas  \emph{Max-v}, \emph{B.EGO} and \emph{Random} are sampling points further away.
Thus, it is not surprising that the trade-off methods, such as \emph{EGO}, as well as the greedy \emph{Max-u} perform well.
Thus, as expected, the distribution of the data is a factor in the relaxation and performance.}

\textcolor{black}{ \subsection{Case study III: Multi-dimensional Curie transition temperature for BaTiO$_3$ based ceramics }}
\textcolor{black}{The Curie transition temperature in BaTiO$_3$ based ceramics is affected by several features or descriptors, and 
 here we test how the transition temperature behaves in terms of multi-dimensional variables without a guiding functional form for the surface.
 In total, we employ 182 pieces of data obtained in our laboratory, 
 and our objective is to decrease the difference between the predicted values and the measurement values.
 The variables are selected based on our previous work that use methods such as gradient boosting with a Kriging based model.
 It has been shown that four features adequately capture the data with the smallest 10-fold CVerror using all combinations of features. 
 For the initial model training we randomly select 30\% of the data.}

\begin{figure}[htbp]
	\begin{center}
		\includegraphics[width = \columnwidth]{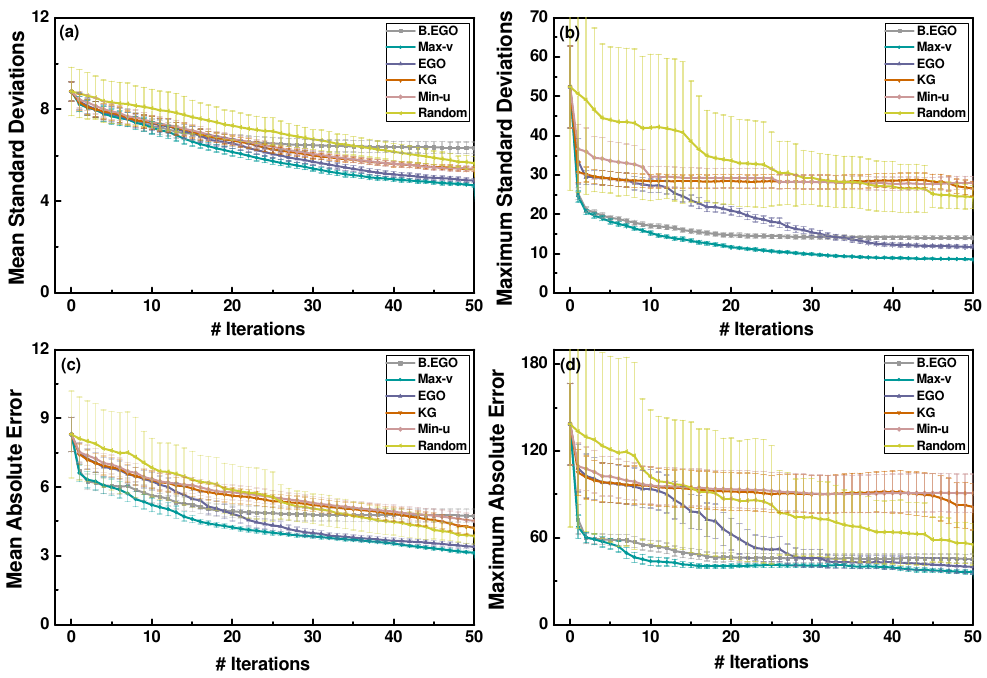}
		\caption{
		\textcolor{black}{Comparison of the performance of utility functions in optimizing the the targeted objective function for the Curie temperature for BaTiO$_3$ based ceramics. The initial data size contains $n$ = 73 training points and  the mean values and  error bars showing the 95\% confidence intervals of the points are evaluated over 100 trials. Shown is the behavior of (a) mean standard deviation,  (b) maximum standard deviation,  (c) mean absolute error, and (d) maximum absolute error.}}
		\label{fig:15}
	\end{center}
\end{figure}

\textcolor{black}{\autoref{fig:15} shows the results  for 100 trials.
There is a significant drop initially in the maximum error for \emph{Max-v} and \emph{B. EGO}, suggesting that the bulk of the uncertainty is reduced within a few iterations.
Again, \emph{Max-v} relaxes the most but the others are not far from converging to it. 
Unlike the other cases,  \emph{Max.AE} does not relax to zero and is indicative of the complexity of the problem. 
We have essentially used only four features to model this data from a sampling of 182 measurement points. 
There are uncertainties in the model itself, hence the Kriging needs to be accurate, and we are assuming that the sampling of points is representative of the data in the whole space.
}
 
 \section{Discussion}
  \begin{figure*}[htbp]
	\begin{center}
		\includegraphics[width = 18cm]{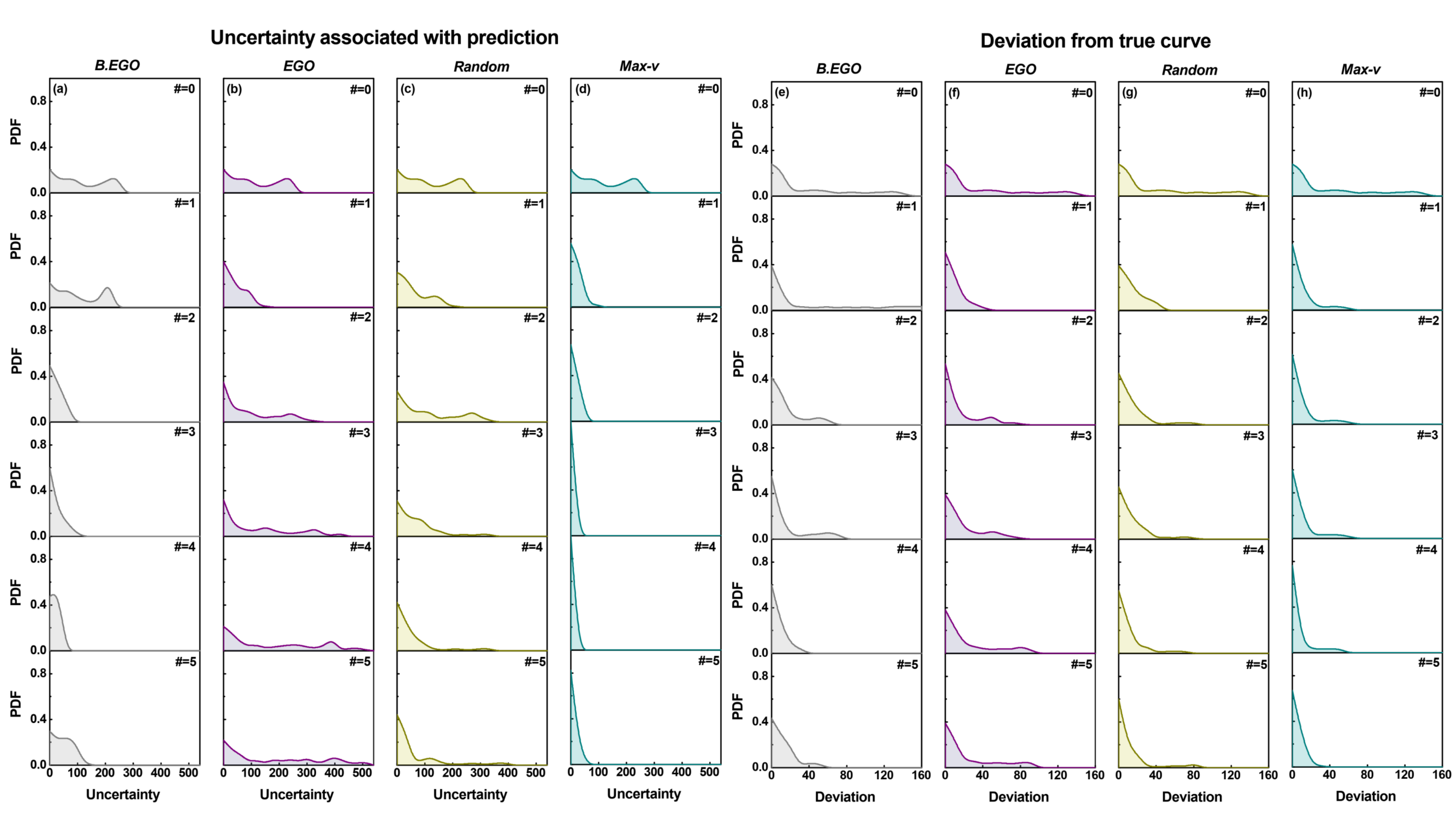}
		\caption{For the  Liquidus line of the Fe-C phase diagram shown in \autoref{fig:7}, the probability density functions (PDF) of the uncertainties and deviation from true result associated with the Kriging model prediction for \emph{Random}, \emph{Max-v}, \emph{B.EGO} and \emph{EGO} are shown with successive iterations. }
		\label{fig:10}
	\end{center}
\end{figure*}

Our objective has been to compare the influence of utility functions for curves, such as phase boundaries, fatigue lines, and other multi-dimensional cases important in materials science. This is essential as in the absence of analytical results, it is difficult to predict {\it a priori} which utilities will be superior in reducing the the costs of acquiring new information when learning from data.

\textcolor{black}{ Except for the random case, all the utilities we compare are based on directed exploration, which can incorporate different degrees of exploitation. 
Our key finding is that maximum variance (\emph{Max-v}) performs very well across a range of data sets with varying complexities, including the addition of experimental noise.  The function, \emph{B.EGO}, which tracks the variability over bootstrap samples, and uses $EGO$ to minimize the variability across the whole function, also shows  relatively good performance, although it is not as robust as  \emph{Max-v}. Moreover, we also find that for each type of data set, there exists a utility which performs as well, if not better than, or at least competes with \emph{Max-v}. The distributions of the property values in the dataset $y$ can also influence the behavior. If the distributions depart from the uniform distribution, then typically there are relatively few training data points located near global minima/maxima,  and  these can be associated with large deviations from the true result. We find this to be the case  for the Hartmann 3 function and intermolecular potential data sets, where the  trade-off methods \emph{EGO} and \emph{KG} perform as well, if not better than  \emph{Max-v}.  
Also, for a given problem, several utilities can converge but at varying iteration numbers.
For the intermolecular potential, the convergence is far superior to  \emph{Max-v} even after 50 iterations.
In cases where  \emph{Random} selection does converge, it requires a lot more iterations as the exploration is unguided.
In the presence of experimental noise, \emph{SKO}, which is essentially \emph{EGO} with noise incorporated, is the superior performer at noise levels of  5\% and 10\%, although {\it Max-v} also does well, converging with more iterations.}

\textcolor{black}{Our results emphasize the importance of  making the appropriate choice in ranking and selecting the next candidate for measurements or calculations.}

\begin{figure*}[htbp]
	\begin{center}
		\includegraphics[width = 18cm]{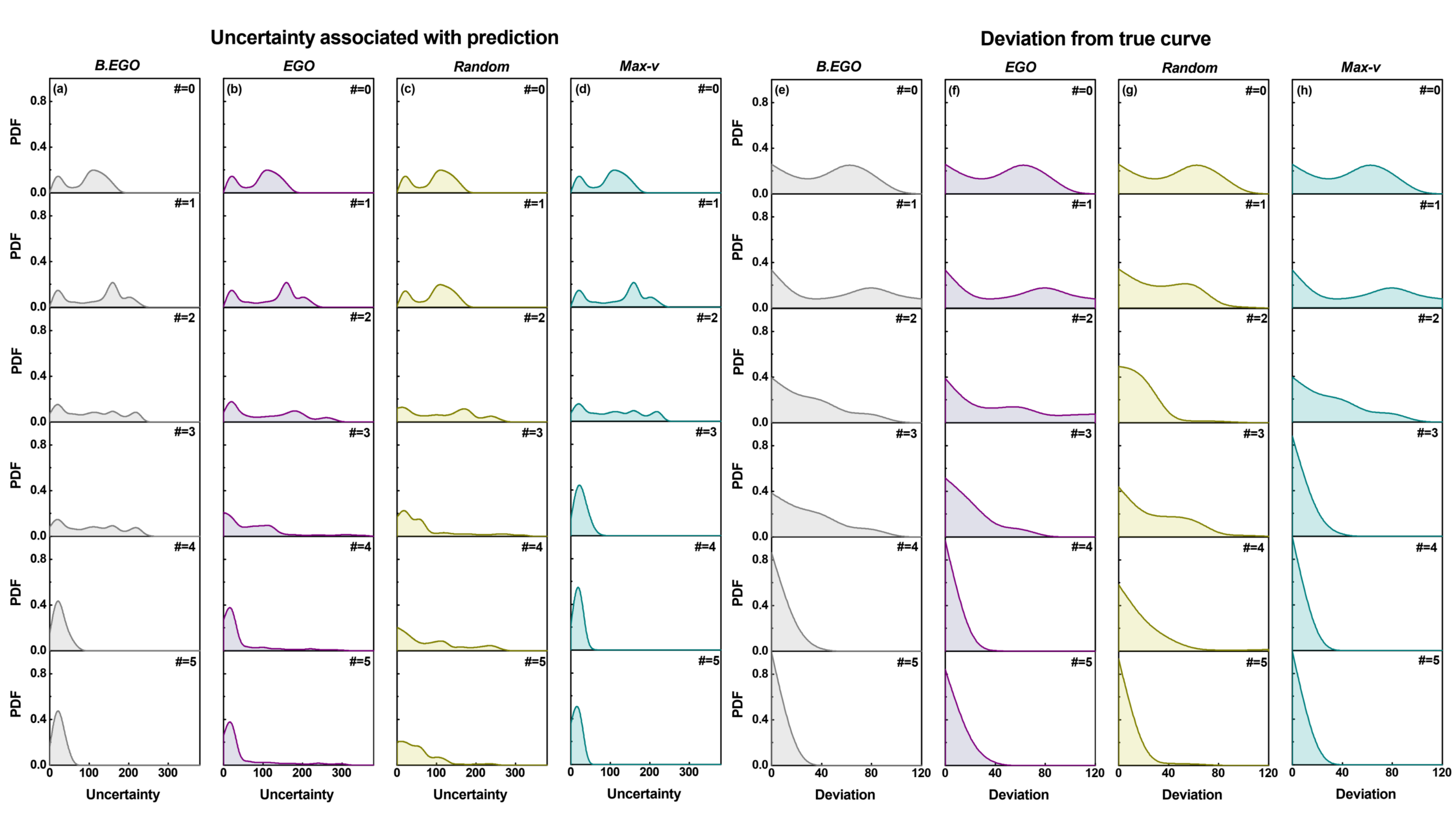}
		\caption{For the Fatigue life curve for SS304L steel shown in \autoref{fig:9},  the probability density functions (PDF) of the uncertainties and deviation from true result associated with the Kriging model prediction for \emph{Random}, \emph{Max-v}, \emph{B.EGO} and \emph{EGO} are shown with successive iterations.}
		\label{fig:11}
	\end{center}
\end{figure*}

To gain an understanding of the behavior of these functions, we 
 plot the probability density functions (PDF) for the uncertainties from the Kriging model estimates and for the deviation of the estimate from the true curve, as a function of iterations. 
\autoref{fig:10} and \autoref{fig:11} show the results for the Liquidus line of Fe-C phase diagram and for the fatigue life curve for 304L steel, respectively.
The Fe-C curve is more complex and its Kriging estimate would not be expected to be as good.
Thus, for all the four utilities being studied, we see wide distributions in  the uncertainty profile for the estimates and the deviation from true curve  (panel \# = 0) of \autoref{fig:10}. 
With successive iterations as the next point is added, the distributions of the uncertainties begin to narrow and the mean value tends towards zero.
All strategies are efficient in this sense with \emph{Max-v} leading to a desired narrower sharply peaked  distribution  (\# = 5) of \autoref{fig:10}.
For \emph{Random} selection and \emph{EGO},  points with very large uncertainties always occur because of the long tails in the distribution.
The other utilities target such points with large uncertainties to reduce the tail in the uncertainty distribution. 

 The Kriging estimate for the fatigue line behavior shown in \autoref{fig:11} is better than for Fe-C as the function by comparison is quite monotonic and so
the initial distributions of ``Deviation from true curve'' shown in  panels (\# = 0) of \autoref{fig:11} are narrower.
The evolution of the distributions for utility functions (except for \emph{Random}) is similar to that in \autoref{fig:10}.
Because the objective function is simpler, all utility functions show very similar performance in the first two iterations.
Thereafter the PDF of ``Deviation from true curve'' employing \emph{Max-v} quickly converges to zero compared to \emph{B.EGO} and \emph{EGO}.
Thus, irrespective of how good initially the predictive model is, \emph{Max-v} shows superior performance in these two 1D cases.

\textcolor{black}{We conclude with some general remarks on circumstances that favor   \emph{Max-v} and trade-off methods such as \emph{EGO}. If the variance is large and the deviation from the true result also large,
then selection by \emph{Max-v} will have a significant affect in decreasing the deviation further.  However, if the deviation is small (i.e. the model is good), then the reduction in deviation will not be significant.  
This is likely the case for both of the 1D curve examples in Case 1 where \emph{Max-v} is especially good. Similarly, in situations where the uncertainty is not too large
, and the deviation from the true result is significant, 
then trade-off methods such as \emph{EGO} will have a substantial effect in locating the max/min of the curve, that is,  decrease the deviation further. We suggest this is the case for the Hartmann 3 function and PES intermolecular potential. Trade-off methods depend on balancing the uncertainty (exploration) with exploitation (model performance), and in the limits where the uncertainties are either very large or very small, for a given deviation, then from expression (4), \emph{EGO} behaves either as  \emph{Max-v} or chooses the  model prediction, respectively. }

\begin{figure}[htbp]
	\begin{center}
		\includegraphics[width = \columnwidth]{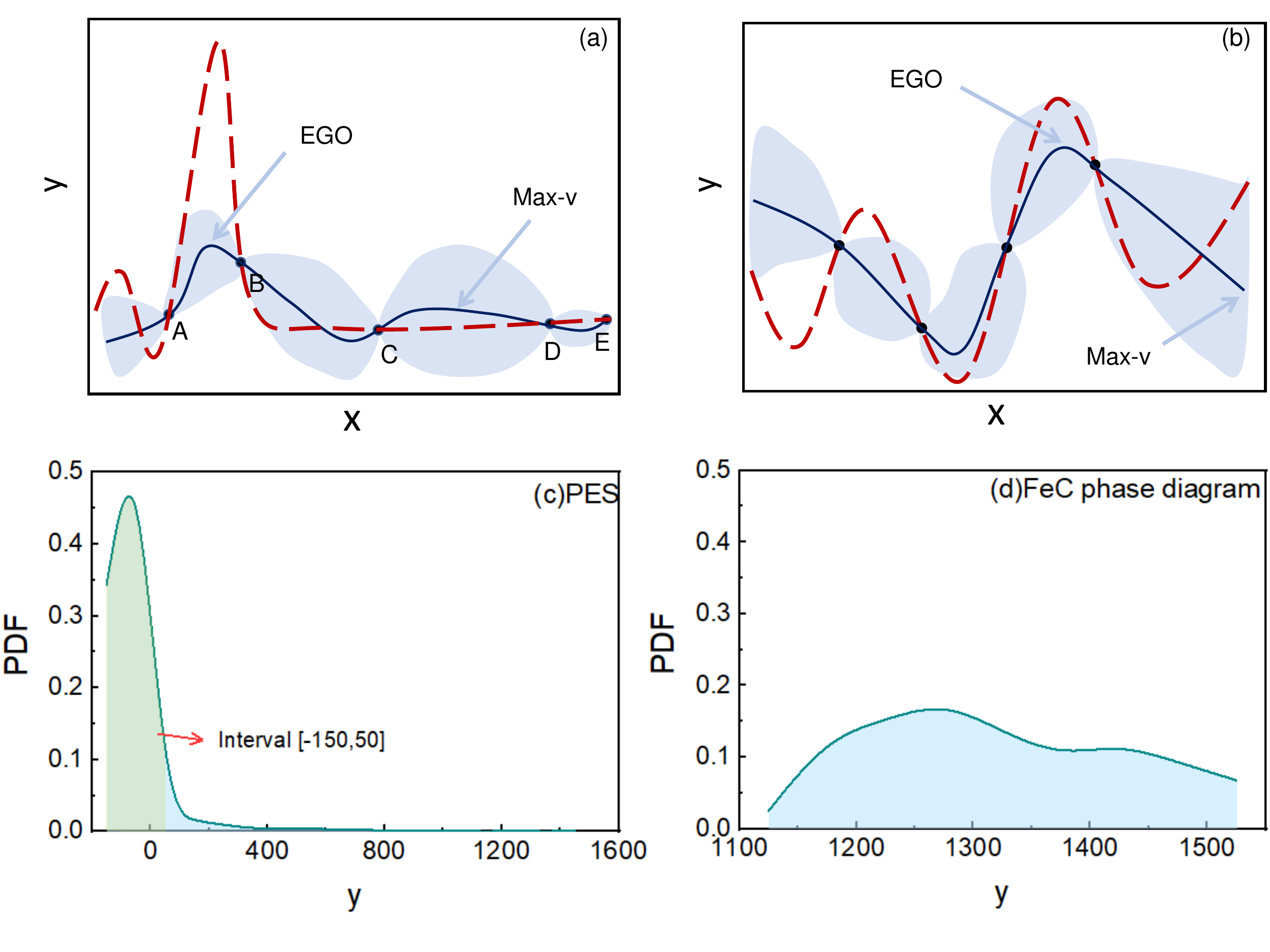}
		\caption{\textcolor{black}{Typical solutions and the probability density functions (PDF) of the property values $y$  for the intermolecular potential energy surface (PES) and the FeC phase boundary datasets. The black line represents the model prediction and the red line the true solution.}}
		\label{fig:19}
	\end{center}
\end{figure}

\textcolor{black}{To illustrate graphically, we have plotted in \autoref{fig:19}(c) and \autoref{fig:19}(d) the distribution of the $y$ data values for the intermolecular potential energy surface (PES) and  FeC phase boundary examples. We note that the former deviates strongly from a uniform distribution of values, whereas the latter is closer to uniform. Schematics of the solutions corresponding to these two cases are shown in  \autoref{fig:19}(a) and \autoref{fig:19}(b), where the red line is the actual solution and the black line the model prediction.  The distribution of $y$ values in \autoref{fig:19}(c) typically gives rise to the curves of  \autoref{fig:19}(a) with a maximum 
and there are relatively few training data points close to the maximum in the distribution. As expected, \emph{EGO}  works well near the maximum, whereas \emph{Max-v} is better where the uncertainties are large. For a more uniform distribution of $y$ data values as in \autoref{fig:19}(d) for the 1D FeC example, the solution profile is more complex as in  \autoref{fig:19}(b), with \emph{Max-v} having a greater impact on driving the optimization towards the solution.
We hope our work will motivate further studies on a variety of materials data to confirm our general findings, as well as provide a deeper understanding of why \emph{Max-v} works so well across data sets with varying complexities.}

\section*{Acknowledgements}
The authors gratefully acknowledge the support of the National Key Research and Development Program of China (2017YFB0702401), and National Natural Science Foundation of China (Grant Nos. 51571156, 51671157, 51621063, and 51931004). 
\section*{References}
\bibliographystyle{naturemag}
\bibliography{selector}
\end{document}